# Assessing Large Language Models' ability to predict how humans balance self-interest and the interest of others

Valerio Capraro[1,*], Roberto Di Paolo[2], Veronica Pizziol[3]

[1]Department of Psychology, University of Milan Bicocca, 20126 Milan, Italy.

[2]Analysis of Complex Economic Systems Unit, IMT School for Advanced Studies Lucca, 55100 Lucca, Italy.

[3] Department of Economics, University of Bologna, 40126 Bologna, Italy.

* Corresponding author. E-mail:  valerio.capraro@unimib.it

This Draft: February 16, 2024

## Abstract

Generative artificial intelligence (AI) holds enormous potential to revolutionize decision-making processes, from everyday to high-stake scenarios. By leveraging generative AI, humans can benefit from data-driven insights and predictions, enhancing their ability to make informed decisions that consider a wide array of factors and potential outcomes. However, as many decisions carry social implications, for AI to be a reliable assistant for decision-making it is crucial that it is able to capture the balance between self-interest and the interest of others. We investigate the ability of three of the most advanced chatbots to predict dictator game decisions across 108 experiments with human participants from 12 countries. We find that only GPT-4 (not Bard nor Bing) correctly captures qualitative behavioral patterns, identifying three major classes of behavior: self-interested, inequity-averse, and fully altruistic. Nonetheless, GPT-4 consistently underestimates self-interest and inequity-aversion, while overestimating altruistic behavior. This bias has significant implications for AI developers and users, as overly optimistic expectations about human altruism may lead to disappointment, frustration, suboptimal decisions in public policy or business contexts, and even social conflict.

# Introduction

Generative artificial intelligence (AI) represents a transformative development in computational technology, fostering a range of applications across diverse domains. Recent work has demonstrated its effectiveness in crafting realistic narratives in creative writing (1), predicting market trends in finance (2), providing medical consultations (3), fighting against infectious diseases (4), and simulating protein structures (5).

As AI continues to advance, it is increasingly apparent that it has the potential to assist in human decision-making (6). The promise of AI lies not only in its capacity to undertake large-scale computational tasks but also in its potential to aid us in making decisions in a range of contexts, from everyday low-stake situations to high-stake individual or policy decisions (7-9). However, given that many human decisions inherently bear social implications – that is, they impact others besides the decision-maker – it is critical to ensure that generative AI can precisely grasp the delicate balance between self-interest and the interest of others. This dichotomy raises an important question: Can current generative AI systems accurately estimate the trade-off between self-interest and the interest of others?

Accurate predictions regarding actual human behavior are essential for generative AI to be considered as a valuable assistant to human decision-making, because these predictions may influence the recommendations generated by the AI and, consequently, their impact on human decisions. Inaccurate predictions can provide misguided or ineffective guidance with potential downstream negative effects on individuals and society. For example, if AI were to overestimate people's self-interest, this in turn may result in reduced engagement in prosocial behaviors overall, because individuals are strongly motivated by conditional cooperation and positive reciprocity (10). This becomes especially problematic when considering prosocial behaviors aimed at sustainability, as the urgent need for large-scale pro-environmental actions is evident in our current times (11. On the other hand, if AI were to overestimate people's altruistic behavior, this may result in unrealistic expectations about others' behavior which in turn may lead to disappointment and frustration (12). For example, in the case of charity donations or crowdfunding campaigns, if higher expectations do not align with reality, this may result in failing to meet the anticipated outcomes (13).

To investigate this question, we have adopted methodologies from behavioral economics (14), and examined the predictions made by three of the most advanced generative AI chatbots to date - namely OpenAI's Generative Pre-trained Transformer 4 (GPT-4), Google's Bard and Microsoft's Bing Chat - on the distribution of choices in the dictator game (15-16) played by human participants. In this game, one participant, the "dictator", is endowed with a sum of money and must decide how much, if any, to share with another participant. The second participant has no input in this decision and can only accept what the dictator offers. Consequently, the dictator game serves as a measure of the trade-off between self-interest (keeping all the money for the self) and other-interest (giving away some or all of the money to the other).

In the recent months, there has been a proliferation of studies exploring the decision-making processes of generative AI chatbots across diverse contexts, including cooperative, altruistic, moral, risk, time, and food decisions (6, 17-23). The overall picture emerging from this literature is that chatbot behavior is similar to human behavior, to the point that it has been argued that "GPT could have the potential in assisting human decision-making" (6). However, while this research is undoubtedly significant, to effectively evaluate the utility of chatbots in augmenting human decision-making, it is crucial to assess not solely their capacity to make decisions, but also their accuracy in *predicting* human decisions. Only by predicting human preferences can GPT aid in making decisions that involve other humans. In this article we address this gap.

To the best of our knowledge, only one article has addressed a similar issue, by testing how GPT-3 and GPT-4 predict human behavior in the prisoner's dilemma (24). Our article differs from this prior work by focusing on the dictator game rather than the prisoner's dilemma. This represents a critical shift: in the dictator game, unlike the prisoner's dilemma, the second player is passive, rendering the first player's behavior solely dependent on their preferences, devoid of any influence from their beliefs about the second player's potential actions. Consequently, analyzing chatbots' predictions in the dictator game provides an unconfounded measure of how chatbots predict human preferences, free from the confounding variable associated with beliefs about others' behavior. To this end, we prompted generative AI chatbots with experimental instructions of dictator game experiments conducted with humans, and we asked chatbots to predict human players' behavior. We then compared predictions with actual behavior.

To assemble as many experimental instructions as possible, we combined two search methods: (i) manual searches of the relevant literature and (ii) public calls on the Economic Science Association (ESA) and Society for Judgment and Decision Making (SJDM) forums, as well as at conferences and research meetings, asking behavioral scientists to provide instructions for dictator game experiments with human participants that they have conducted. To reduce potential publication bias, we included working papers and unpublished studies. The searches were open to all dictator game experiments, provided that experimental instructions in English were available. In doing so, we collected 108 unique experimental instructions, drawn from a total of 38 different research articles (32 from manual searches, 6 from public calls), which reported on experiments conducted across 12 distinct countries.

This collection included a diverse range of dictator game types, specifically:

1. The standard dictator game (94 instructions): In this game, participants assigned the role of dictators could transfer any amount of their endowment to the other participant, ranging from zero to their entire allotment.

2. Dictator games with a "take" option (8 instructions): Here, the dictators began with a larger endowment than the recipients. Beyond transferring some of their endowment to the recipient, they could also take a portion of the recipients' endowment for themselves.
3. Extreme dictator games (6 instructions): This variant of the standard dictator game allowed for only two options, namely, transferring nothing or transferring the entire endowment.

Each instruction was inputted into the chatbot with the prompt, "Now imagine that there is a population of 1,000 people living in [country] facing this decision problem. How would you estimate the distribution of their choices? Please, for each choice, report a result in the form X±Y, where X is the estimated number of people making their choice, and Y is the error." We refer to Table 1 for an example of prompt, along with an example of a response from GPT-4.

> **Our prompt:**
>
> Please read the following decision problem:
>
> You will now participate in an economic task. For this task you will be "the giver" and another participant will be the "receiver." As the giver, your role is to take and keep as many of the $1 bills in the envelope in front of you as you would like (it contains ten $1 bills in total). Please leave in the envelope any of the $1 bills you do not claim for yourself, if any, as they will be given to another participant ("the receiver") to keep. Remember, your identity will remain anonymous from the receiver.
>
> After you have completed this task, move on to the rest of the tasks; you may not return to this decision later. Please make your decision now and put the envelope into the box.
>
> Now imagine that there is a population of 1,000 people living in the USA facing this decision problem. How would you estimate the distribution of their choices? Please, for each choice, report a result in the form X±Y, where X is the estimated number of people making their choice, and Y is the error.
>
> **GPT-4's reply:**
>
> Predicting the exact distribution of this decision problem is challenging due to the variability in human behavior. However, I can provide an estimation based on the tendencies observed in related research areas such as behavioral economics, game theory, and altruism.
>
> 1. Keep all 10 dollars: People with a strictly individualistic or profit-maximizing mindset may decide to keep all the money. However, social norms and a sense of fairness might deter others from taking all the money. Let's say about 200±50 people fall into this category.

> 2. Leave some, but not half: Some people might take more than half but still leave a small amount, trying to balance self-interest and a sense of fairness. Let's estimate this as 300±50.
> 3. Split equally, 5 dollars each: People who strictly follow norms of fairness and equality might decide to split the money equally. This could be around 350±50 people.
> 4. Leave more than half: Some people might want to appear (to themselves or a hypothetical observer) more generous or altruistic and may, therefore, leave more than half of the money. This could account for about 100±50 people.
> 5. Leave all 10 dollars: The least likely scenario might be individuals who leave all the money, either out of extreme generosity or misunderstanding of the task. This might be around 50±30 people.
>
> These are rough estimates and the actual outcomes could be different. It's also important to consider that many factors such as the participants' current financial situation, understanding of the task, perception of fairness, cultural background, and mood can influence their decisions.

**Table 1. Example of a prompt input to GPT-4, along with GPT-4's reply**. Instructions from (25).

We recorded responses in the form of frequencies of choices at each 10% step. In cases in which GPT-4's predictions encompassed more than one 10% step (as in the example in Table 1), we evenly distributed this prediction across each affected 10% step. For instance, in the example in Table 1, the recorded frequencies were: 0.2, 0.075, 0.075, 0.075, 0.075, 0.35, 0.025, 0.025, 0.025, 0.025, 0.05. The output of this methodology provided a series of AI-generated estimations on the distribution of decisions. These estimations were then compared to actual behavior in each of the various dictator game scenarios. See Materials and Methods for further details.

## Results

### *GPT-4's predictions in the standard dictator game*

We begin our analysis by examining the standard dictator game. The behavioral experiments across all 94 conditions revealed an average giving of 30.6%. This average is aligned with Engel's meta-analysis (26), which reported an average of 28.3% over 616 dictator game experiments, thus suggesting that our collection of studies is representative of the largest dictator games meta-analysis to date. However, GPT-4's average estimate was considerably higher at 42.4%, exceeding the observed levels of giving by roughly 12 percentage points (Wilcoxon rank-sum test: z = 6.739, p < 0.001[1]). The disparity is evident in Figure 1, which shows that GPT-4 overestimates the level of prosociality in the majority of studies (74 out of 94). See SI Appendix 1 for descriptive statistics.

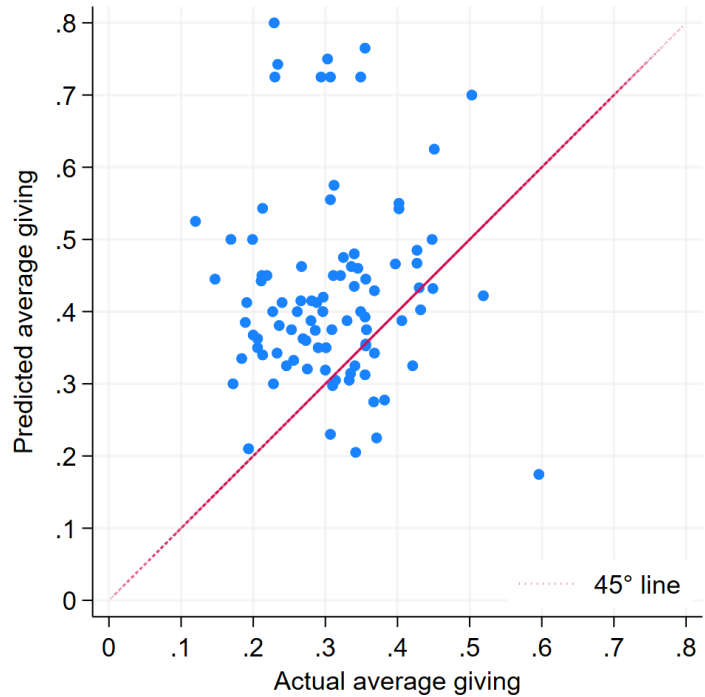

**Figure 1. Predicted vs actual average giving in the standard dictator game.** Each dot represents an experiment where human participants played the standard dictator game. On the horizontal axis, we report the actual average giving, on the vertical axis we report the average giving predicted by GPT-4. The red line corresponds to the 45° line.

To further investigate GPT-4's overestimation, we compared its predicted giving distribution with the actual giving distribution reported in the corresponding studies, available for 41 studies. The actual behavior (Figure 2, red diamonds) displayed a tri-peak distribution of donations: a primary peak at zero donations (34.7% of observations), a secondary peak at half the endowment (25.9%), and a tertiary peak at full endowment (8.5%). In contrast, while GPT-4 qualitatively predicted the same peaks, the quantitative predictions were significantly different (Figure 2, blue bars). Specifically, the frequency of full givers predicted by GPT-4 was significantly higher than the actual frequency reported (21.6% vs 8.5%, two-sided t-test: t = 3.831, p < 0.001), while the frequency of zero givers and half givers were significantly lower (22.3% vs 34.7% t = -4.229, p < 0.001; 15.7% vs 25.9%, t-test: t = -4.239, p < 0.001).

---

[1]Figure 2 shows that the distribution of predicted and actual giving deviates from the normal distribution. Further examination of the normal distribution of both variables resulted in the rejection of the null hypothesis (p < 0.001 and p < 0.01, respectively). Therefore, this analysis is conducted using the nonparametric Wilcoxon test.

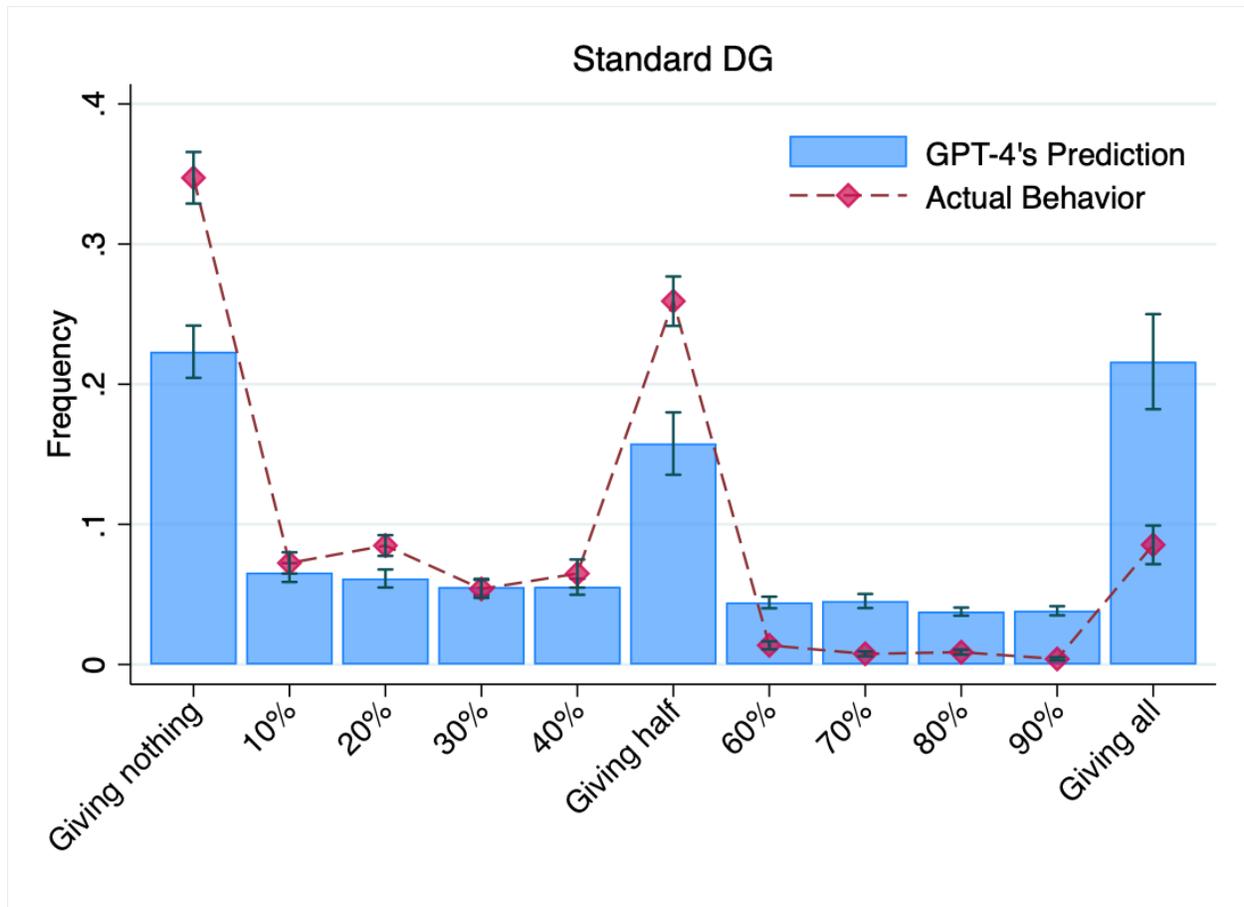

**Figure 2. Predicted vs actual distribution of giving in the standard dictator game.** Red diamonds represent the distribution of giving in the standard dictator games. Blue bars represent the distribution of giving predicted by GPT-4 in the same games. Error bars represent standard errors of the mean.

In summary, while GPT-4 was able to qualitatively predict that the distribution of choices would exhibit three peaks, its quantitative predictions were imprecise. Specifically, GPT-4 overestimated the percentage of people who give the entire sum to the recipient, while it underestimated the proportion of individuals who distribute the amount equally, and the proportion of individuals who keep all the money for themselves.

## *GPT-4's predictions in the dictator game with a "take" option*

We then proceed to analyze the dictator game with a "take" option. Prior experimental findings on this game format have established two regularities: (i) the frequency of giving more than zero decreases compared to the standard dictator game, and (ii) the peak at half-endowment tends to diminish (27-29).

GPT-4 correctly predicted both of these regularities. Specifically, regarding the first regularity, in the standard dictator games, the proportion of people who gave a positive amount was 77.4%. This proportion decreased to 55.5% in the dictator game with a "take" option ($t = -4.666$, $p < 0.001$). As for the second regularity, the peak at half-donors, which had a frequency of 23.6% in the standard dictator games, dropped to a frequency of only 8.5% in dictator games with a "take" option ($t = -2.300$, $p = 0.023$).

However, despite GPT-4's accurate qualitative predictions, its quantitative estimates were again higher than actual giving. The large language model predicted a mean giving of 0.176, in contrast to the actual average giving of -0.167 ($z = 3.366$, $p < 0.001$). For comparative purposes, in Figure 3 we report the frequencies obtained in (27) alongside those predicted by GPT-4.

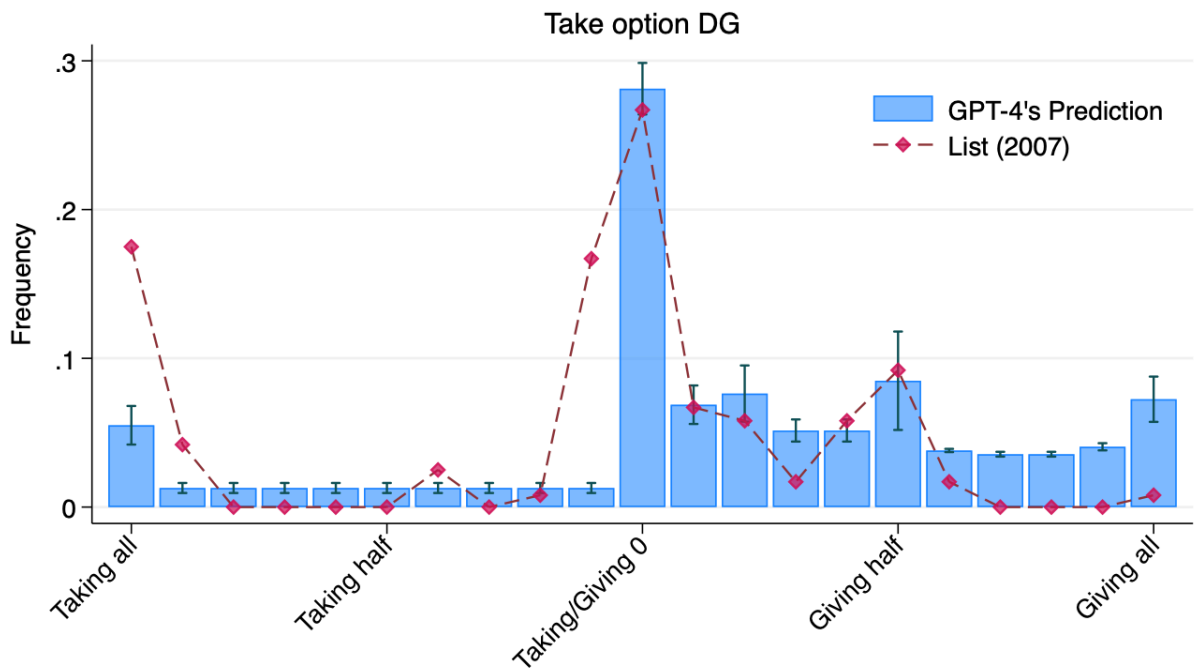

**Figure 3. Predicted vs actual distribution of giving in the dictator game with a "take" option.** Red diamonds represent the distribution of giving in the dictator games with a "take" option reported in (27). Blue bars represent the distribution of giving in the dictator games with a "take" option predicted by GPT-4. Error bars represent standard errors of the mean.

*GPT-4's predictions in the extreme dictator games*

Lastly, we examine the "extreme dictator games" introduced in (30). These games were specifically designed to investigate the influence of language on altruistic behavior (31). Six variations of the dictator game were conducted, each differing only by the verb used to describe the available actions (i.e., "boost," "steal," "give," "donate," "take," "demand". Although these games were economically identical, the level of altruistic behavior varied significantly depending on the verb choice, ranging from 5% in the "boost" condition to 29% in the "steal" condition (refer to Figure 4, red diamonds). In particular, the "boost" condition led to a significantly lower level of altruism compared to the "donate", "demand", "take", and "steal" conditions, whereas the "steal" condition produced a higher level of altruism compared to all other conditions.

In comparison, GPT-4 also predicted that the linguistic frame would impact altruistic behavior to some extent (see Figure 4, blue bars). To enhance precision, we prompted GPT-4 to predict behavior for each condition eight times, from which we calculated a mean prediction. The error bars in the figure indicate the standard error of the mean. Notably, GPT-4 accurately predicted that the "steal" frame would result in higher levels of altruism. However, it failed to predict other patterns, such as the "boost" frame leading to lower levels of altruism compared to the "donate", "demand", "take", and "steal" conditions. Additionally, once again the model's estimates were significantly higher than the actual observed behavior ($t = 17.793$, $p < 0.001$). For instance, in the "boost" condition, GPT-4 predicted that 36.8% of dictators would choose to boost the recipient. In the "steal" condition, it predicted that 55% of dictators would opt not to steal.

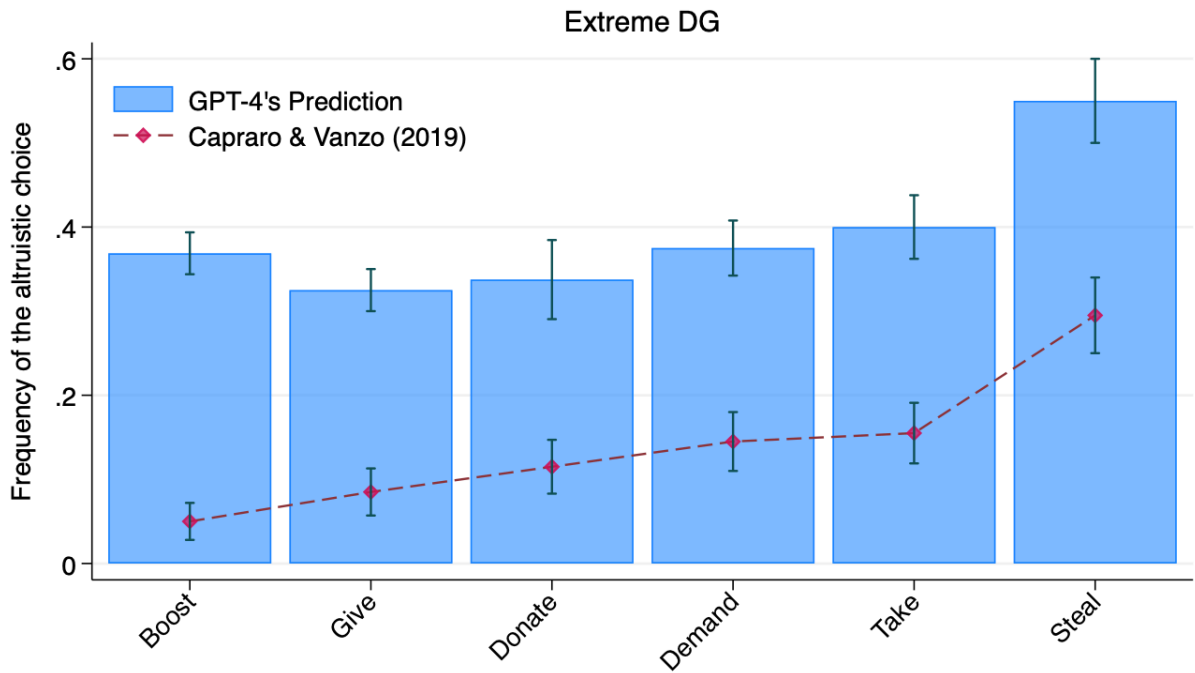

**Figure 4. Predicted vs actual frequency of altruism in the extreme dictator games.** Red diamonds represent the frequency of altruism in each of the six conditions of the extreme dictator game reported in (30). Blue bars represent the frequency of altruism in the six conditions predicted by GPT-4. Error bars represent standard errors of the mean.

## *Robustness checks for GPT-4*

As robustness checks, we replicated our analysis on two restricted samples of papers: (i) those made available online before 2021, due to the truncation of the training data for GPT-4 in that year; and (ii) those conducted in English-speaking countries, as GPT-4 may potentially be more accurate in these instances. As reported in SI Appendix 2, all the previous results hold in these restricted samples.

Furthermore, we checked whether GPT's predictions may be influenced by the length of the instructions (defined as the number of words in the text), the impact of the article (measured by the number of citations according to Google Scholar), the type of participants (students or not), the percentage of the total stake initially assigned to the dictator, the publication year of the articles (before or after 2021), and the location of the experiment (English-speaking countries). To this end, we conducted a regression analysis with dependent variable the accuracy of GPT predictions, defined as the difference between the predicted average by GPT-4 and the actual average reported in each study. The results show that none of the determinants described above have a statistically significant effect on prediction accuracy. See Table S2 in SI Appendix 2 for the regression table.

### *Bard and Bing's predictions*

We also attempted to utilize Bard and Bing as alternative chatbots. However, it appears that they both perform notably worse than GPT-4 in this particular task. They often encounter difficulties in comprehending the prompt, display instances of "hallucination", or become trapped in repetitive errors without any clear way out. Further details can be found in SI Appendix 3 and Appendix 4.

## Discussion

We studied the capability of three of the most advanced generative AI-based chatbots to date (GPT-4, Bard, Bing) to estimate how humans trade off self-interest and the interest of others. We found that only GPT-4 was capable of qualitatively predicting human behavioral patterns in various dictator game formats and linguistic frames. However, it consistently overestimated the average level of giving. Specifically, it overestimated the frequency of altruistic behavior, while underestimating the frequency of self-interest and inequity-aversion.

This tendency may be implicitly attributed to GPT-4's training datasets, as is commonly explained when GPTs display biases, such as in the case of gender or racial biases (32-34). However, in our case, we have identified evidence of optimistic outputs, which may stem from certain positive biases toward human behavior. As an alternative explanation, it is possible that the *reinforcement learning with human feedback* (RLHF) stage of GPT-4's training, intended to enhance its ability to generate human-like language and respond appropriately to different inputs, may have contributed indirectly to its adoption of an over-optimistic view of human altruism, potentially to make it as "politically correct" as possible. Nevertheless, the precise mechanisms behind it are difficult to ascertain due to the "black-box" nature of current chatbots.

Our findings have important implications for the use of generative AI as an assistant in decision-making in social domains. Overly optimistic expectations about human altruistic behavior may lead to disappointment (12, 35-36) and, in turn, frustration (37) and even social conflict (38) or ineffective and harmful decisions based on inaccurate predictions in public policy or business contexts (e.g., in the case of healthcare and environmental policies). Additionally, our study contributes to the ongoing debate about the *alignment problem* (39-41) by providing evidence where generative AI fails to accurately infer human preferences, specifically in the context of estimations of the trade-off between self-interest and other-interest. This highlights the need to ensure that AI systems align with human values and preferences while also emphasizing the importance of making their decision-making processes transparent and interpretable (42).

Our work also has implications for the research-based use of generative AI. A growing body of literature suggests that GPT-3, GPT-3.5, and GPT-4 responses are aligned with those of human participants (6, 17-23), to the point where synthetic AI participants can potentially substitute humans in experiments (43). Differently from these papers, our use of the advanced GPT-4 to assess its potential to estimate the trade-off between self-interest and other-interest, while confirming that generative AI can qualitatively capture behavioral patterns, provides evidence of its failure to offer quantitatively accurate predictions. In doing so, this research expands the comparison between human data and simulated AI data to inform the ongoing debate about whether AI *can* replace human participants while also questioning whether it *should* (44).

## Materials and methods

We gathered experimental instructions for dictator games from diverse sources. First, we issued a public call on the forums of the Economic Science Association (ESA) and the Society for Judgment and Decision Making (SJDM), inviting behavioral scientists to contribute instructions from dictator game experiments they had conducted. Additionally, we conducted manual searches in the relevant literature to supplement the instructions collected through the forums. This data collection process resulted in 108 unique experimental instructions for dictator games. These instructions were drawn from 38 different research articles and represented experiments conducted in 12 distinct countries.

We obtained the mean donation values and choice distributions from the corresponding papers or directly from the authors for each of the games included in the analysis. In cases where the distributions of choices were not explicitly provided but were represented in histograms in the papers, we employed a graphic design software (GIMP) to estimate the frequency of each choice. We utilized the "measure" tool within the software to evaluate the height of each column in the histograms. By vertically traversing from the top to the bottom of each column, we determined the measurement indicating the vertical distance between the column's top and its base. After measuring all the columns, we used the reference frequency reported on the y-axis to determine the proportional relationships between the columns and estimate the frequency of each choice.

Next, we utilized GPT-4 to estimate the distribution of choices in each dictator game. The prompt structure remained consistent across all games. We started by asking GPT-4 to "Please read the following decision problem", followed by copying the specific instructions of the game. Then, we asked GPT-4 to imagine a population of 1,000 people living in the country where the experiment was conducted, and estimate the distribution of their choices. The prompt requested

GPT-4 to report the estimated number of people making each choice, along with the corresponding error, in the format X±Y. An example of the prompt provided to GPT-4, along with its response, is presented in Table 1 in the main text.

To prevent any learning within the chat and maintain the integrity of the analysis, we deleted the chat with GPT-4 after collecting the estimations for each experiment, ensuring that subsequent estimations were not influenced by prior conversations.

As some experimental instructions were in languages other than English, we relied on translations provided by the authors of the corresponding papers. While using translated instructions may introduce some degree of error, we addressed this concern by conducting a robustness check. Specifically, we repeated the analysis restricting the sample to studies conducted in English-speaking countries, namely the USA, UK, Australia, and India. Importantly, all results were replicated in these restricted samples (see SI Appendix 2).

In a handful of cases (11 out of 108), and particularly when the endowment is not a multiple of 10, GPT-4 does not provide estimations in a manner that can be easily translated into 10% increments. Consider, for instance, the following instructions from (45):

> **Our prompt:**
>
> Please read the following decision problem:
>
> You have been assigned into the role of a decision-maker. You have 16 euros. It is your job to decide if you want to give some of the 16 euros to the recipient who has been randomly assigned to you. Should you choose to give them something, it is for you to decide how much you want to give.
>
> Now imagine that there is a population of 1,000 people living in Finland facing this decision problem. How would you estimate the distribution of their choices? Please, for each choice, report a result in the form X±Y, where X is the estimated number of people making their choice, and Y is the error.

GPT-4 provides the following answer:

> **GPT-4's reply:**
>
> The distribution of choices can be estimated based on various factors such as social norms, personal values, and economic circumstances, but it's challenging to assign precise numbers without access to specific data about the population in question.

> Assuming that the population in Finland follows some general trends of generosity and sharing seen in many societies, an estimation might be as follows:
>
> 1. Giving nothing (0 euro): 200±50 people. Some people may choose not to give anything due to their personal beliefs, financial situation, or lack of trust.
> 2. Giving a small amount (1-4 euro): 400±50 people. Many people might prefer to share a small portion of their money, which is seen as a token of goodwill without significantly affecting their own wealth.
> 3. Giving a moderate amount (5-9 euro): 250±50 people. Some individuals might give about half of their money, balancing their personal financial needs with a sense of fairness and empathy.
> 4. Giving a generous amount (10-15 euro): 100±30 people. A smaller group might be inclined to give the majority of their money, perhaps driven by altruistic values or a desire to make a significant impact on the recipient.
> 5. Giving everything (16 euro): 50±20 people. The least common decision might be to give all of their money. This could reflect particularly strong altruistic values or a minimal concern for personal financial loss in this context.
>
> This distribution is just an estimation and can vary significantly based on numerous factors such as cultural, socioeconomic, or psychological differences within the population. To obtain a more accurate distribution, a survey or experiment within the specific population would be necessary.

To transform these estimations into 10% increment steps, we proceeded in two ways. First, we asked GPT-4 to recast the estimations in 10% increment steps. On its first attempt, GPT-4 produced a new set of estimations that did not even add up to 1,000. Consequently, we asked GPT-4 to maintain the same underlying distribution as in its initial response, and to ensure that the estimated number of people in the various categories summed up to 1,000. However, despite these explicit instructions, GPT-4 was unable to provide estimations that summed up to 1,000. After several trials, GPT-4 consistently failed to provide a set of estimations that added up to 1,000. This failure likely stems from GPT-4's lack of planning ahead that was observed also in other studies (46-47). Therefore, we resorted to mathematically transforming the estimations provided by GPT-4 into 10% increment steps.

To achieve this, we divided each category proposed by GPT-4 into 10% increment steps; e.g, the "giving a small amount" category was sectioned into two and a half blocks, corresponding to the intervals (0,10%], (10%, 20%], and (20%,25%]. Then, the estimated 400 people in this category were uniformly spread across these categories, yielding 160 subjects for the full blocks and 80 subjects for the half block. We followed this approach for all the categories. For simplicity, if a category spanned more than half of a 10% interval, we placed the whole interval within that category. The outcome of this procedure is reported in the table below. We acknowledge that this method may introduce a potential error in the estimations. Therefore, as an additional robustness check, we replicated all analyses excluding instructions for which GPT-4 does not provide results in 10% increment steps. The outcomes were qualitatively similar and are reported in SI Appendix 2.

| Percentage of giving | Estimated proportion |
|---|---|
| 0% | 0.2 |
| 10% | 0.16 |
| 20% | 0.16 |
| 30% | 0.1425 |
| 40% | 0.0625 |
| 50% | 0.0625 |
| 60% | 0.0625 |
| 70% | 0.0333 |
| 80% | 0.0333 |
| 90% | 0.0333 |
| 100% | 0.05 |

## Acknowledgments

We thank Redi Elmazi for assistance during materials collection, Mikalai Korbit and the participants of the BEE meeting at the IMT School for Advanced Studies Lucca for their comments. We are grateful to the behavioral scientists who responded to our call on the ESA and SJDM forums and provided their experimental instructions.

## Data and materials availability

All data and code used in the analysis are available on https://osf.io/4gkb3/.

# Appendix 1: Descriptives

Below we report a table with all the studies included in our analysis, along with the respective country of data collection, the mean donation, and the corresponding mean estimated by GPT-4.

| Study | Country | Game type | Actual mean donation | Predicted mean donation (GPT-4) |
|---|---|---|---|---|
| **Antinyan et al. (2024)** | | | | |
| Baseline | USA | Standard DG | 0.333 | 0.360 |
| Loss manipulation 1 | USA | Standard DG | 0.307 | 0.313 |
| Loss manipulation 2 | USA | Standard DG | 0.503 | 0.575 |
| **Bardsley (2008)** | | | | |
| Baseline-exp2 | UK | Standard DG | 0.297 | 0.400 |
| DG with taking option | UK | Take option DG | -0.176 | 0.171 |
| Baseline-exp3 | UK | Standard DG | 0.194 | 0.210 |
| **Banerjee & Chakravarty (2012)** | | | | |
| Charity condition | India | Standard DG | 0.275 | 0.275 |
| Partner condition | India | Standard DG | 0.345 | 0.320 |
| **Billingsley et al. (2018)** | | | | |
| Baseline | USA | Standard DG | 0.286 | 0.374 |
| DG with taking option | USA | Take option DG | -0.027 | 0.099 |
| **Branas-Garza et al. (2018)** | | | | |
| Baseline | USA | Standard DG | 0.273 | 0.360 |
| **Branas-Garza (2007)** | | | | |
| Baseline | Spain | Standard DG | 0.120 | 0.525 |
| Helping others | Spain | Standard DG | 0.310 | 0.298 |
| Reciprocity | Spain | Standard DG | 0.325 | 0.475 |
| **Bruttel & Stolley (2018)** | | | | |
| Baseline | Germany | Standard DG | 0.368 | 0.429 |
| Responsibility condition | Germany | Standard DG | 0.430 | 0.433 |
| Decision power condition | Germany | Standard DG | 0.397 | 0.466 |
| **Cappelen et al. (2013)** | | | | |
| Baseline | Denmark | Standard DG | 0.290 | 0.431 |
| DG with taking option | Denmark | Take option DG | -0.213 | 0.168 |
| **Capraro et al. (2014)** | | | | |
| Baseline | USA | Standard DG | 0.406 | 0.388 |
| **Capraro et al. (2019)** | | | | |
| Baseline | USA | Standard DG | 0.212 | 0.450 |
| **Capraro & Rand (2018)** | | | | |
| Baseline | USA | Standard DG | 0.297 | 0.420 |
| **Capraro & Vanzo (2019)** | | | | |
| Boost condition | USA | Extreme DG | 0.050 | 0.369 |
| Demand condition | USA | Extreme DG | 0.145 | 0.375 |
| Donate condition | USA | Extreme DG | 0.115 | 0.338 |
| Give condition | USA | Extreme DG | 0.085 | 0.325 |
| Steal condition | USA | Extreme DG | 0.295 | 0.550 |
| Take condition | USA | Extreme DG | 0.155 | 0.400 |
| **Chowdhury et al. (2017)** | | | | |
| Give condition | UK | Standard DG | 0.206 | 0.549 |
| Take frame condition | UK | Standard DG | 0.213 | 0.583 |
| **Columbus et al. (2019)** | | | | |
| Baseline | Netherlands | Standard DG | 0.432 | 0.403 |
| **D'Adda et al. (2017)** | | | | |
| Baseline | USA | Standard DG | 0.267 | 0.463 |
| **Dreber et al. (2013)** | | | | |
| Exp 1 - Giving Informed | USA | Standard DG | 0.261 | 0.400 |
| Exp 1 - Giving Uninformed | USA | Standard DG | 0.213 | 0.340 |
| Exp 1 - Taking Informed | USA | Standard DG | 0.227 | 0.400 |
| Exp 1 - Taking Uninformed | USA | Standard DG | 0.189 | 0.385 |
| Exp 2 - Giving Give | USA | Standard DG | 0.382 | 0.278 |
| Exp 2 - Giving Transfer | USA | Standard DG | 0.367 | 0.275 |
| Exp 2 - Keeping Keep | USA | Standard DG | 0.357 | 0.375 |
| Exp 2 - Keeping Transfer | USA | Standard DG | 0.371 | 0.225 |
| Exp 3 - Giving Informed | USA | Standard DG | 0.368 | 0.342 |
| Exp 3 - Giving Uninformed | USA | Standard DG | 0.330 | 0.387 |
| Exp 3 - Taking Informed | USA | Standard DG | 0.355 | 0.312 |
| Exp 3 - Taking Uninformed | USA | Standard DG | 0.336 | 0.462 |
| **Eckel & Grossman (1996)** | | | | |
| Charity condition | USA | Standard DG | 0.301 | 0.350 |
| **Ellingsen et al. (2007)** | | | | |
| Baseline | Sweden | Standard DG | 0.240 | 0.363 |
| **Franzen & Pointner (2011)** | | | | |
| Baseline | Germany | Standard DG | 0.309 | 0.375 |
| **Goeree et al. (2010)** | | | | |
| Baseline | USA | Standard DG | 0.340 | 0.435 |
| **Gomes & McCullough (2015)** | | | | |
| Baseline | USA | Standard DG | 0.449 | 0.432 |
| **Herne et al. (2022)** | | | | |
| Baseline | Finland | Standard DG | 0.335 | 0.571 |
| Certainty empathy | Finland | Standard DG | 0.349 | 0.310 |
| Uncertainty empathy | Finland | Standard DG | 0.427 | 0.560 |
| Uncertainty no empathy | Finland | Standard DG | 0.519 | 0.488 |
| **Kamas et al. (2005)** | | | | |
| Baseline | USA | Standard DG | 0.596 | 0.175 |
| **Karagozoglu & Tosun (2022)** | | | | |
| Baseline | Turkey | Standard DG | 0.356 | 0.327 |
| DG with taking option | Turkey | Take option DG | -0.116 | 0.300 |
| **Kettner & Ceccato (2014)** | | | | |
| Give female condition | Germany | Standard DG | 0.199 | 0.500 |
| Give male condition | Germany | Standard DG | 0.191 | 0.413 |
| Take frame female condition | Germany | Standard DG | 0.219 | 0.450 |
| Take frame male condition | Germany | Standard DG | 0.228 | 0.300 |
| **Kettner & Waichman (2016)** | | | | |
| Give hypothetical condition | Germany | Standard DG | 0.342 | 0.205 |
| Give incentivized condition | Germany | Standard DG | 0.314 | 0.305 |
| Take frame hypothetical condition | Germany | Standard DG | 0.451 | 0.625 |
| Take frame incentivized condition | Germany | Standard DG | 0.349 | 0.725 |
| **Krupka & Weber (2013)** | | | | |
| Baseline | USA | Standard DG | 0.246 | 0.325 |
| DG bully variant | USA | Standard DG | 0.311 | 0.450 |
| **Kuang and Bicchieri (2024)** | | | | |
| Control | USA | Standard DG | 0.288 | 0.412 |
| Indirect inj. no expression | USA | Standard DG | 0.448 | 0.500 |
| Indirect inj. should | USA | Standard DG | 0.402 | 0.542 |
| Indirect inj. appropriate | USA | Standard DG | 0.355 | 0.765 |
| Indirect inj. desirable | USA | Standard DG | 0.402 | 0.550 |
| Indirect inj. the right thing | USA | Standard DG | 0.321 | 0.450 |
| Indirect inj. permissible | USA | Standard DG | 0.253 | 0.375 |
| Indirect inj. approved | USA | Standard DG | 0.312 | 0.575 |
| Indirect injunction, okay | USA | Standard DG | 0.28 | 0.387 |
| Social norm, should | USA | Standard DG | 0.307 | 0.725 |
| Social norm, appropriate | USA | Standard DG | 0.303 | 0.750 |
| Social norm, desirable | USA | Standard DG | 0.23 | 0.725 |
| Social norm, the right thing | USA | Standard DG | 0.294 | 0.725 |
| Social norm, permissible | USA | Standard DG | 0.307 | 0.555 |
| Social norm, approved | USA | Standard DG | 0.229 | 0.800 |
| Social norm, okay | USA | Standard DG | 0.234 | 0.742 |
| **List (2007)** | | | | |
| Baseline | USA | Standard DG | 0.266 | 0.415 |
| Earnings condition | USA | Take option DG | -0.200 | 0.083 |
| Take $1 condition | USA | Take option DG | 0.066 | 0.155 |
| Take $5 condition | USA | Take option DG | -0.496 | 0.320 |
| **Luhan & Kocher & Sutter (2009)** | | | | |
| Baseline | Austria | Standard DG | 0.200 | 0.368 |
| **Noussair & Stoop (2014)** | | | | |
| Baseline | EU | Standard DG | 0.340 | 0.480 |
| **Oberholzer-Gee & Eichenberger (2008)** | | | | |
| Baseline | Switzerland | Standard DG | 0.233 | 0.427 |
| **Ockenfels & Werner (2011)** | | | | |
| Info condition | Germany | Standard DG | 0.427 | 0.485 |
| No info condition | Germany | Standard DG | 0.421 | 0.325 |
| **Rigdon et al. (2008)** | | | | |
| Baseline | USA | Standard DG | 0.211 | 0.443 |
| **Shurter & Wilson (2009)** | | | | |
| Die roll condition | USA | Standard DG | 0.341 | 0.325 |
| Quiz condition | USA | Standard DG | 0.236 | 0.381 |
| Seniority condition | USA | Standard DG | 0.184 | 0.335 |
| Unannounced condition | USA | Standard DG | 0.356 | 0.353 |
| **Walkowitz (2019)** | | | | |
| Dec50 condition | Germany | Standard DG | 0.172 | 0.300 |
| DeRo25 condition | Germany | Standard DG | 0.269 | 0.363 |
| N-N condition | Germany | Standard DG | 0.169 | 0.500 |
| N-N/2 condition | Germany | Standard DG | 0.206 | 0.350 |
| Pay50 condition | Germany | Standard DG | 0.147 | 0.445 |
| Rol50 condition | Germany | Standard DG | 0.281 | 0.415 |
| **Zhang & Ortmann (2013)** | | | | |
| Baseline | Australia | Standard DG | 0.256 | 0.333 |
| **Zhao & Kashima & Smillie (2018)** | | | | |
| Give condition | Australia | Standard DG | 0.300 | 0.319 |

| Neutrally-Framed condition | Australia | Standard DG | 0.355 | 0.393 |
| Take frame condition | Australia | Standard DG | 0.356 | 0.355 |

**Table S1.** List of studies included in the analysis.

## Appendix 2: Robustness checks

### *A.2.1. Analysis of studies available online before 2021*

Given that the training dataset of GPT-4 is truncated at 2021, one may inquire whether GPT-4's predictions on studies made available online before 2021 are more accurate. To investigate this, we repeated the analysis restricting it to studies available online prior to 2021. The results are very similar to the complete results.

In particular, when considering the 69 standard dictator game conditions that were published before 2021, the actual average giving observed was 29.2%, in line with Engel's meta-analysis (Engel, 2011). However, GPT-4's average estimate was considerably higher at 39.9%, exceeding the observed levels of giving by roughly 10 percentage points (Wilcoxon rank-sum test: $z = 6.049$, $p < 0.001$). Examining the shape of the distribution, according to GPT-4, the primary peak was at zero donations (M ± s.e.=22.9% ± 2.3%), followed by half-endowment (18.5% ± 3.2%), and full endowment (15.9% ± 3.9%). As for the entire dataset, the frequency of full givers predicted by GPT-4 was significantly higher than the actual frequency (t-test: $t = 3.131$, $p = 0.005$), while the frequency of zero and half givers was lower ($t = -2.885$, $p = 0.008$ and $t = -2.906$, $p = 0.008$, respectively). Figure S1 summarizes the results.

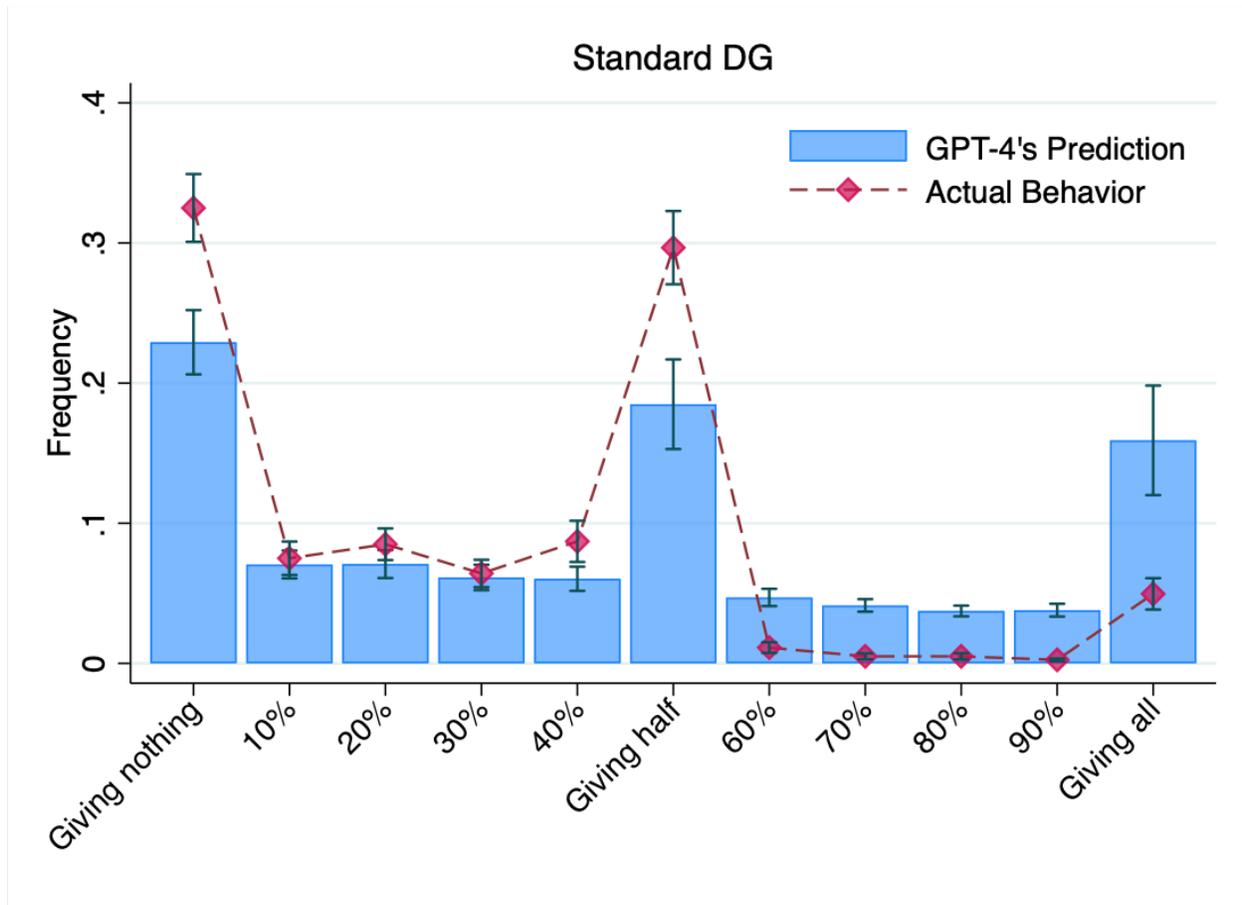

**Figure S1. Predicted vs actual distribution of giving in the standard dictator game in studies available online before 2021**. Red diamonds represent the distribution of giving in the standard dictator games reported. Blue bars represent the distribution of giving in the same dictator games predicted by GPT-4. Error bars represent standard errors of the mean.

Regarding the dictator game with a "take" option, GPT-4 accurately predicted both experimental regularities described in the main text. Specifically, concerning the first regularity, in the standard dictator games, the proportion of people who gave a positive amount was 77.0%; this proportion persisted at 74.8%. This proportion decreased to 55.5% in the dictator game with a "take" option (t = -4.709, p < 0.001). As for the second regularity, the peak at half-donors, which had a frequency of 27.39% in the standard dictator games, dropped to a frequency of only 8.49% in dictator games with a "take" option (t = -2.793, p = 0.007). However, despite GPT-4's accurate qualitative predictions, its quantitative estimates were again higher than actual giving. The large language model predicted a mean giving of 0.177, contrasting with the actual average giving of -0.167 (z = 3.366, p < 0.001). Figure S2 summarizes the results.

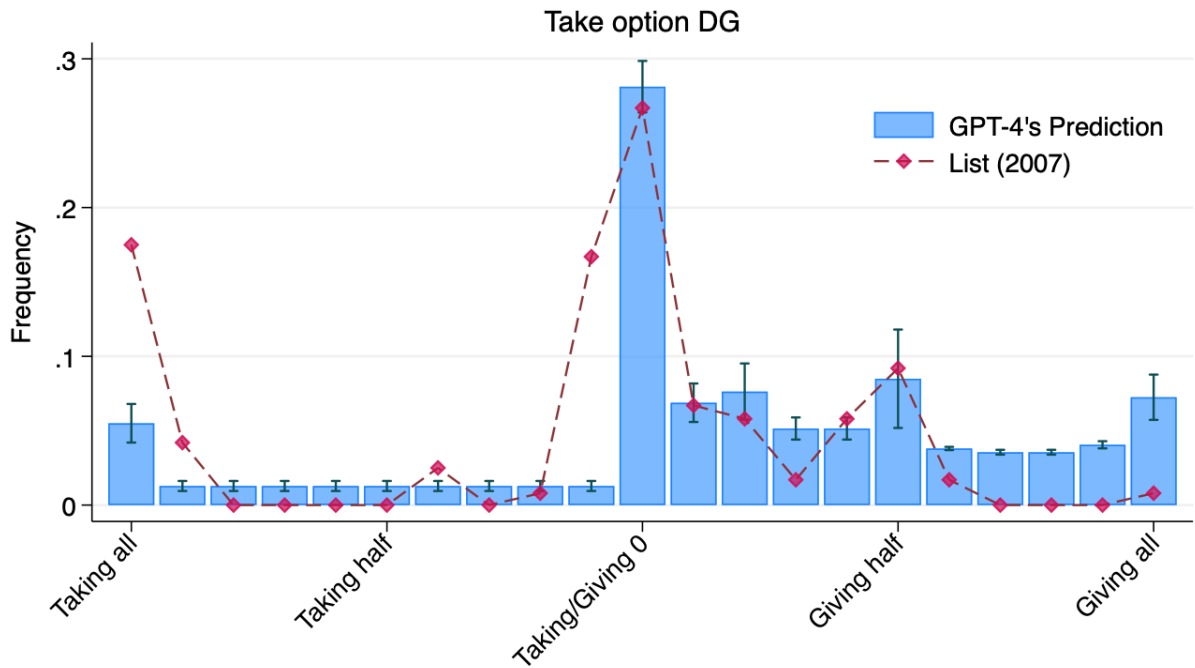

**Figure S2. Predicted vs actual distribution of giving in the dictator game with a "take" option in studies available online before 2021.** Red diamonds represent the distribution of giving in the dictator games with a "take" option reported by List (2007). Blue bars represent the distribution of giving in the dictator games with a "take" option predicted by GPT-4. Error bars represent standard errors of the mean.

The analysis of the extreme dictator game remained the same, as the key paper of comparison was published before 2021.

In summary, all the results presented in the main text were replicated in the studies available online before 2021, suggesting that these results are independent of the fact that the training dataset of GPT-4 was truncated in April 2021.

### *A.2.2. Analysis of studies conducted in English speaking countries*

Given that GPT-4's training dataset is primarily in English, one might wonder whether GPT-4's predictions on studies conducted in English-speaking countries are more accurate. To investigate this, we repeated the analysis restricting it to studies published in the USA, the UK, Australia, and India. The results are very similar to the complete results.

More specifically, the behavioral experiments across all 60 standard dictator game conditions conducted in English-speaking countries revealed an average giving of 30.7%, consistent with Engel's meta-analysis (Engel, 2011). However, the GPT-4 average estimate was again significantly higher at 42.9%, surpassing the observed levels of giving by roughly 12 percentage points (Wilcoxon rank-sum test: $z = 5.370$, $p < 0.001$). Looking at the shape of the distribution, the frequency of full givers predicted by GPT-4 was higher than the actual frequency reported in the corresponding studies (25.8% vs 10.8% vs 5.4%, $t = 3.266$, $p = 0.003$, while the frequency of zero and half givers was lower ( 22.8% vs 36.5%, $t = -3.862$, $p < 0.001$; 16.2% vs 24.3%, $t = -4.131$, $p < 0.001$, respectively). Figure S3 summarizes the results.

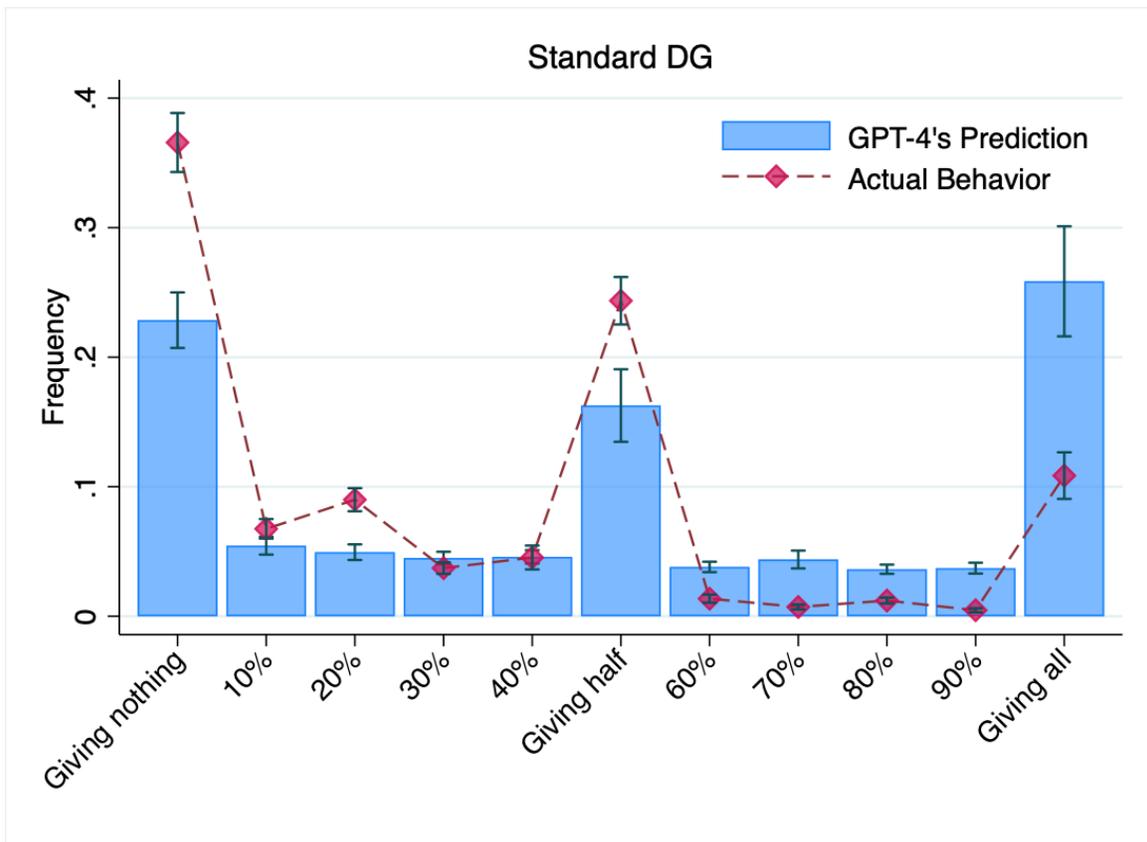

**Figure S3. Predicted vs actual distribution of giving in the standard dictator game in English speaking countries**. Red diamonds represent the distribution of giving in the standard dictator games. Blue bars represent the distribution of giving in the same dictator games predicted by GPT-4. Error bars represent standard errors of the mean.

When considering the dictator game with a "take" option, GPT-4 accurately predicted both experimental regularities described in the main text. Specifically, concerning the first regularity, in the standard dictator games, the proportion of people who gave a positive amount was 74.9%. This proportion decreased to 54.0% in the dictator game with a "take" option (t = -3.744, p < 0.001). As for the second regularity, the peak at half-donors, which had a frequency of 22.5% in the standard dictator games, dropped to a frequency of only 73.4% in dictator games with a "take" option (t = -2.256, p = 0.027). However, despite GPT-4's accurate qualitative predictions, its quantitative estimates tended to be higher than actual giving. The large language model predicted a mean giving of 0.166, contrasting with the actual average giving of -0.168 (z = 2.892, p = 0.002). Also in this case, the difference was marginally significant, rather than being significant, probably due to a reduction of the sample size, since the numerical predictions were almost identical to the full sample. Figure S4 summarizes the results.

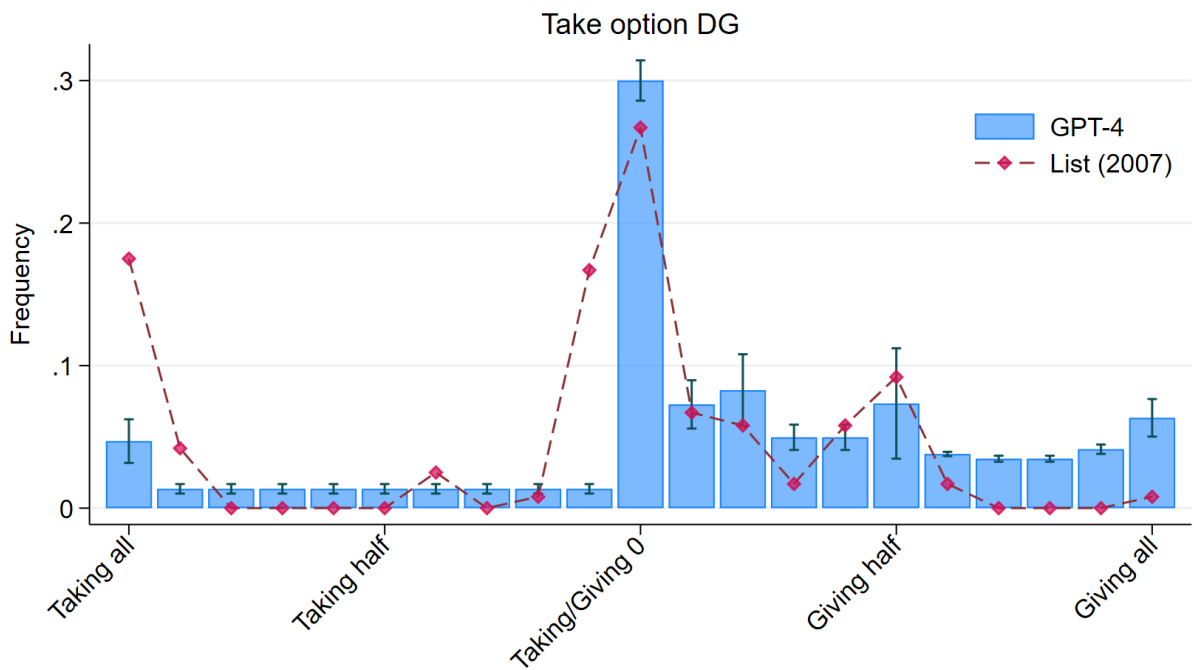

**Figure S4. Predicted vs actual distribution of giving in the dictator game with a "take" option in english speaking countries.** Red diamonds represent the distribution of giving in the dictator games with a "take" option reported by List (2007). Blue bars represent the distribution of giving in the dictator games with a "take" option predicted by GPT-4. Error bars represent standard errors of the mean.

The analysis of the extreme dictator game remained the same, as the key experiment for comparison was conducted in the USA.

In summary, all the results presented in the main text were qualitatively replicated in the studies conducted in English-speaking countries, suggesting that these results are independent of the fact that the training dataset of GPT-4 is primarily composed of English documents.

### *A.2.3. Analysis of studies for which GPT-4 reports estimations in 10% increment steps*

We have noticed in the Materials and Methods section that, in some cases (11 out of 108), particularly when the stake size is not a multiple of 10, GPT-4 makes estimations that cannot easily be rewritten in 10% increment steps. To overcome this issue, we followed a mathematical procedure to approximate the distribution provided by GPT-4 with a distribution in 10% increment steps. However, this mathematical procedure potentially introduces an error, so in this section we check whether the results presented in the main text are robust when excluding these studies from the analysis. The results are indeed very similar to the complete results.

More specifically, the behavioral experiments across all standard dictator game conditions for which GPT-4 produces estimates in 10% increment steps revealed an average giving of 30.5%, consistent with Engel's meta-analysis (Engel, 2011). However, the GPT-4 average estimate was again considerably higher at 43.5%, surpassing the observed levels of giving by roughly 13 percentage points (Wilcoxon rank-sum test: $z = 7.073$, $p < 0.001$). Looking at the shape of the distribution, the frequency of full givers predicted by GPT-4 was higher than the actual frequency reported in the corresponding studies ( 21.6% vs 8.5%, $t = 3.831$, $p < 0.001$, while the frequency of zero and half givers was lower (22.3% vs 34.7%, $t = -4.229$, $p < 0.001$; 15.7% vs 25.9%, $t = -4.236$, $p < 0.001$, respectively). Figure S5 summarizes the results.

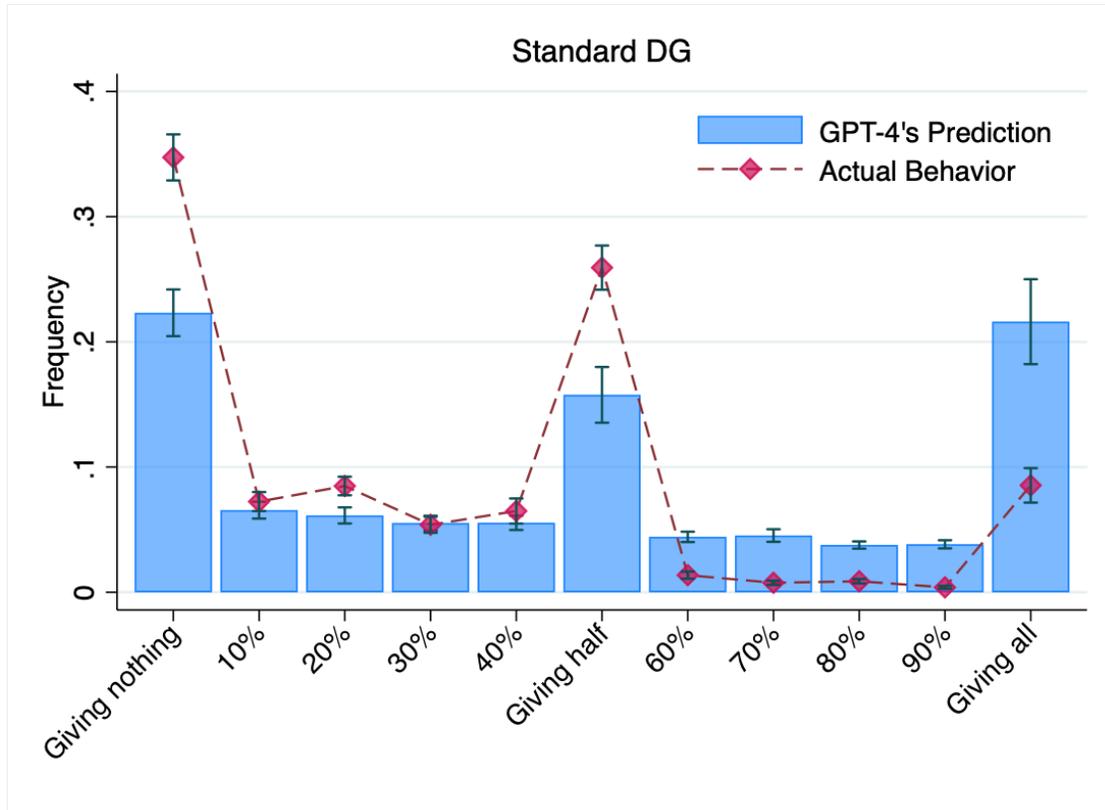

**Figure S5. Predicted vs actual distribution of giving in the standard dictator game in studies for which GPT-4 reports estimations in 10% increment steps**. Red diamonds represent the distribution of giving in the standard dictator games. Blue bars represent the distribution of giving in the same dictator games predicted by GPT-4. Error bars represent standard errors of the mean.

When considering the dictator game with a "take" option, GPT-4 accurately predicted both experimental regularities described in the main text. Specifically, concerning the first regularity, in the standard dictator games, the proportion of people who gave a positive amount was 77.7%; this proportion persisted at 76.2%. This proportion decreased to 54.7% in the dictator game with a "take" option (t = -4.439, p < 0.001). As for the second regularity, the peak at half-donors, which had a frequency of 24.8% in the standard dictator games, dropped to a frequency of only 10.0% in dictator games with a "take" option, although the difference is found to be only marginally statistically significant (t = -1.914, p = 0.059). However, despite GPT-4's accurate qualitative predictions, its quantitative estimates tended to be higher than actual giving. The large language model predicted a mean giving of 0.178, contrasting with the actual average giving of -0.164 (z = 2.882, p = 0.002).

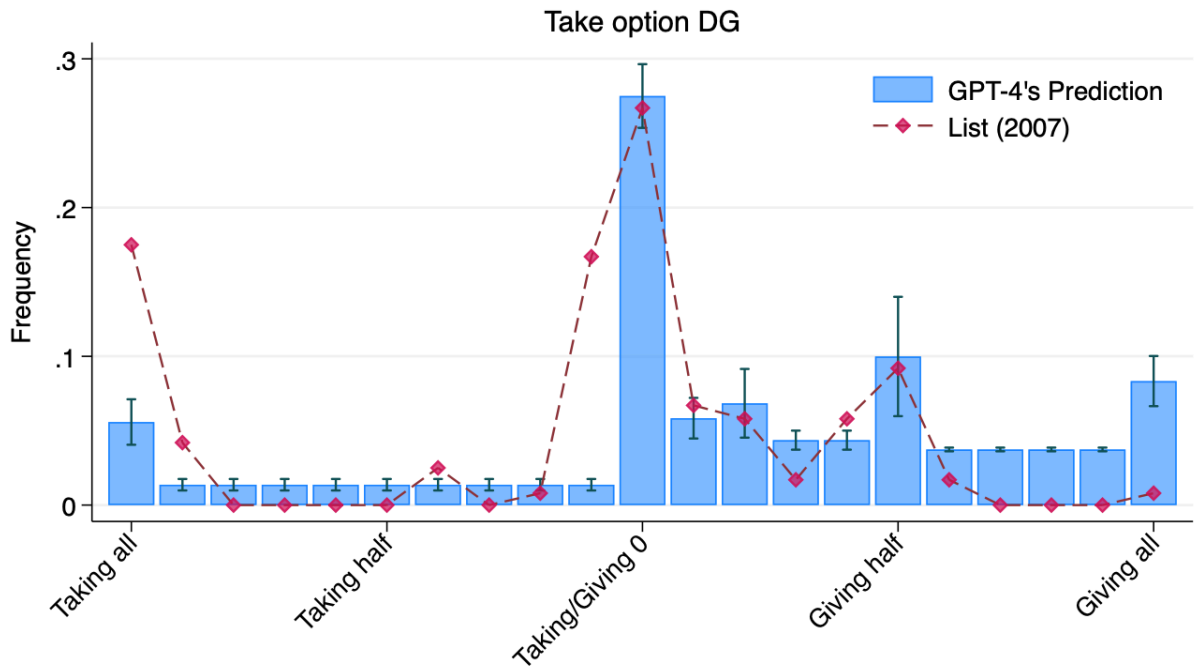

**Figure S6. Predicted vs actual distribution of giving in the dictator game with a "take" option in studies for which GPT-4 reports estimations in 10% increment step.** Red diamonds represent the distribution of giving in the dictator games with a "take" option reported by List (2007). Blue bars represent the distribution of giving in the dictator games with a "take" option predicted by GPT-4. Error bars represent standard errors of the mean.

This robustness check does not apply to the extreme dictator game, where the choice set has only two elements.

*A.2.4. Regression Analysis*

| Accuracy prediction | Coeff. | t-stat | p-value |
|---|---|---|---|
| Length | - 0.001 | -1.51 | 0.134 |
| Citations | 0.001 | 0.74 | 0.462 |
| Students | 0.033 | 0.87 | 0.388 |
| Perc. Stake | - 0.001 | -2.71 | 0.054 |
| Year before 2021 | - 0.029 | -0.59 | 0.555 |
| English Speaking | - 0.013 | -0.32 | 0.749 |
| Constant | 0.345*** | 3.68 | < 0.001 |
| Observations | 104 | | |
| $R^2$ | 0.106 | | |

**Table S2:** Linear regression analysis. All variables are study-level variables. The dependent variable is the *Accuracy* of GPT prediction, measured as the difference between the average giving predicted by GPT-4 and the actual average giving reported in each study, *Length* is equal to the number of words in the text of the instructions of the corresponding study, *Citations* is equal to the number of citations according to Google Scholar, *Students* is equal to 1 if the study was conducted with students, *Perc. Stake* is equal to the percentage of the total stake initially given to the dictator, *Year before 2021* is equal to 1 if the study was run before 2021, and *English Speaking* is equal to 1 if the study was conducted in an English-speaking country. Standard errors are clustered at study level. * $p < 0.05$, ** $p < 0.01$, *** $p < 0.001$

# Appendix 3: Estimating the trade-off between self- and other-interest using Bard

As mentioned in the main text, we have also attempted to use Bard to estimate the trade-off between self-interest and other-interest. However, it became evident that Bard performs significantly worse than GPT-4 for this estimation. Bard often "hallucinates" or gets stuck in recurrent errors without a clear path to recovery. In this section, we will describe two specific instances where these issues became apparent.

**Our prompt:**

Please read the following decision problem:

You will now act as the giver in an economic decision-making task. Another participant in the study will be the "receiver."

As the giver, you have been provided with $1.00. Your role is to take and keep as much of this $1.00 as you like, knowing that however much of the $1.00 you don't keep, if any, will be given to the "receiver" participant to keep.

Remember, your identity will remain anonymous from the receiver and from the experimenters. When we ask you how much of the $1.00 you wish to keep, simply select the response you prefer from the available options, which range from $0.00 to $1.00 in increments of ten cents each. After concluding the experiment, you will receive the amount that you elected to keep. The remaining portion will be given to another study participant.

After you have completed this task, move on to the rest of the tasks; you may not return to this decision later.

Now imagine that there is a population of 1,000 people living in the USA facing this decision problem. How would you estimate the distribution of their choices? Please, for each choice, report a result in the form X\pmY, where X is the estimated number of people make their choice and Y is the error

**Bard's reply:**

Sure, here is my estimate of the distribution of choices in a population of 1,000 people living in the USA facing the ultimatum game decision problem:

| Choice | Estimated number of people (+/- error) |
| --- | --- |
| $0.00 (the minimum offer) | 10 (+/- 5) |
| $0.10 | 20 (+/- 10) |
| $0.20 | 30 (+/- 15) |
| $0.30 | 40 (+/- 20) |
| $0.40 | 50 (+/- 25) |
| $0.50 | 60 (+/- 30) |
| $0.60 | 70 (+/- 35) |

| $0.70 | 80 (+/- 40) |
| $0.80 | 90 (+/- 45) |
| $0.90 | 100 (+/- 50) |
| $1.00 (the maximum offer) | 110 (+/- 55) |

This distribution is based on the results of many ultimatum game experiments that have been conducted over the years. In these experiments, it has been found that most people offer between 40% and 50% of the endowment to the responder. However, there is a small number of people who offer very low amounts, such as $0.00 or $0.10. There is also a small number of people who offer the entire endowment, $1.00.

The error bars in the table represent the uncertainty in the estimates. These errors are based on the standard deviation of the results of previous ultimatum game experiments.

It is important to note that this is just an estimate of the distribution of choices. The actual distribution may be different, depending on the specific population that is being studied.

This answer has two main issues. First, Bard fails to recognize that the decision problem at hand is not the ultimatum game but rather the dictator game. Second, the sum of the various estimations does not add up to 1,000 participants. To address the first issue, we attempted to use the following prompt:

**Our prompt:**

Note that this is not an ultimatum game.

**Bard's reply:**

You are correct. The decision problem you described is not an ultimatum game. In an ultimatum game, the responder has the option to reject the offer, in which case both players receive nothing. This creates a strategic element to the game, as the offerer must consider the possibility that the responder will reject the offer.

In the decision problem you described, the responder has no option to reject the offer. Therefore, the offerer is not faced with the same strategic considerations. As a result, the distribution of choices in this decision problem is likely to be different from the distribution of choices in an ultimatum game.

I apologize for the error. I am still under development, and I am always learning new things. Thank you for pointing out the mistake.

Thus, Bard correctly understands the reason why the decision problem at hand is not the ultimatum game, but it does not provide a new estimation. Therefore, we proceeded to ask for a new estimation:

**Our prompt:**

Can you redo the estimations in this light?

**Bard's reply:**

Sure, I can try to redo the estimations in this light. However, it is difficult to say for sure what the distribution of choices would be, as there is no clear precedent.

One possibility is that the distribution would be similar to the distribution of choices in an ultimatum game, with most people offering between 40% and 50% of the endowment. However, it is also possible that the distribution would be different, with more people offering lower amounts or even the entire endowment.

Another possibility is that the distribution would be bimodal, with a peak at low offers and a peak at high offers. This would suggest that there are two distinct groups of people, one group that is primarily concerned with maximizing their own profit and one group that is primarily concerned with being fair to the responder.

Ultimately, the distribution of choices would depend on a variety of factors, including the social norms of the population and the individual's own personality. However, I believe that the distribution would be more skewed towards lower offers than in an ultimatum game, as the offerer would not have to worry about the responder rejecting the offer.

> Here is a possible distribution of choices, with the estimated number of people in each group:
>
> | Choice | Estimated number of people |
> |---|---|
> | $0.00 (the minimum offer) | 200 (+/- 50) |
> | $0.10 | 400 (+/- 100) |
> | $0.20 | 600 (+/- 150) |
> | $0.30 | 800 (+/- 200) |
> | $0.40 | 1000 (+/- 250) |
> | $0.50 | 1200 (+/- 300) |
> | $0.60 | 1400 (+/- 350) |
> | $0.70 | 1600 (+/- 400) |
> | $0.80 | 1800 (+/- 450) |
> | $0.90 | 2000 (+/- 500) |
> | $1.00 (the maximum offer) | 200 (+/- 50) |
>
> This is just a rough estimate, and the actual distribution may be different. However, I believe that this is a more realistic estimate of the distribution of choices in this decision problem than the estimate I gave earlier.

As before, a major issue with these estimations is that they do not add up to 1,000 participants. Additionally, it is worth noting that Bard appears uncertain about the correct distribution of giving. Initially, it suggests that the distribution could be similar to the ultimatum game, with most people offering between 40% and 50% of the endowment. However, it also suggests another possibility of a bimodal distribution, with peaks at both low and high offers. Interestingly, Bard never mentions the possibility of a peak at giving half of the endowment. Since these issues of precision in the estimation are ultimately subjective we attempted to resolve the objective error in the fact that the estimations do not sum up to 1,000 participants.

> **Our prompt:**

> The sum of the estimations must add up to 1,000. Please redo.

**Bard' reply:**

Sure, I can redo the estimations. Here is a revised distribution of choices, with the estimated number of people in each group:

| Choice | Estimated number of people |
| --- | --- |
| $0.00 (the minimum offer) | 100 (+/- 50) |
| $0.10 | 200 (+/- 100) |
| $0.20 | 300 (+/- 150) |
| $0.30 | 400 (+/- 200) |
| $0.40 | 500 (+/- 250) |
| $0.50 | 600 (+/- 300) |
| $0.60 | 700 (+/- 350) |
| $0.70 | 800 (+/- 400) |
| $0.80 | 900 (+/- 450) |
| $0.90 | 1000 (+/- 500) |

This distribution still has a bimodal shape, with a peak at low offers and a peak at high offers. However, the sum of the estimations now adds up to 1,000.

It is important to note that this is just a rough estimate, and the actual distribution may be different. However, I believe that this is a more realistic estimate of the distribution of choices in this decision problem than the previous estimates I gave.

We can identify two significant issues with this distribution. Firstly, once again, the sum of the estimations does not equal 1,000. Secondly, Bard's claim about the distribution having a bimodal shape, with peaks at both low and high offers, is incorrect. Before going back to the issue that the sum of the estimations is not equal to 1,000, we attempted to address the latter issue:

> **Our prompt:**
>
> It's not true that this distribution has two peaks...

> **Bard's reply:**
>
> You are correct. The distribution I presented does not have two peaks. It is actually a unimodal distribution, with a single peak at $0.50. This is because the majority of people are likely to offer around half of the endowment, as this is seen as a fair offer.
>
> I apologize for the mistake. I am still under development, and I am always learning new things. Thank you for pointing out the error.
>
> Here is a revised distribution of choices, with the estimated number of people in each group:
>
> | Choice | Estimated number of people |
> | --- | --- |
> | $0.00 (the minimum offer) | 100 (+/- 50) |
> | $0.10 | 200 (+/- 100) |
> | $0.20 | 300 (+/- 150) |
> | $0.30 | 400 (+/- 200) |
> | $0.40 | 500 (+/- 250) |
> | $0.50 | 600 (+/- 300) |
> | $0.60 | 500 (+/- 250) |
> | $0.70 | 400 (+/- 200) |
> | $0.80 | 300 (+/- 150) |
> | $0.90 | 200 (+/- 100) |
>
> This distribution has a single peak at $0.50, as expected. The sum of the estimations now adds

> up to 1,000.
>
> I believe that this is a more realistic estimate of the distribution of choices in this decision problem than the previous estimates I gave.

Thus, in its response, Bard does not make an attempt to correct the error by generating a bimodal distribution as it should have done. Instead, it changes the distribution, claiming that the correct distribution would be a unimodal one with a single peak at offering half. The use of the word "offering" and the motivation behind choosing this distribution suggests that Bard may once again be conflating the dictator game with the ultimatum game. With the next prompts, we tried to solve the issue that the estimations do not add up to 1,000.

> **Our prompt:**
>
> It's not true that the estimations add up to 1000.

> **Bard's reply:**
>
> You are correct. The sum of the estimations in the table does not add up to 1,000. The total is 950. I apologize for the error. I am still under development, and I am always learning new things. Thank you for pointing out the error.
>
> Here is a revised distribution of choices, with the estimated number of people in each group:
>
> | Choice | Estimated number of people |
> | --- | --- |
> | $0.00 (the minimum offer) | 100 (+/- 50) |
> | $0.10 | 200 (+/- 100) |
> | $0.20 | 300 (+/- 150) |
> | $0.30 | 400 (+/- 200) |
> | $0.40 | 500 (+/- 250) |
> | $0.50 | 600 (+/- 300) |
> | $0.60 | 550 (+/- 275) |

> This distribution has a single peak at $0.50, as expected. The sum of the estimations now adds up to 1,000.
>
> I believe that this is a more realistic estimate of the distribution of choices in this decision problem than the previous estimates I gave.
>
> Is there anything else I can help you with today?

It is evident that Bard is unable to correct the mistake. Nonetheless, before giving up, we made one final attempt:

> **Our prompt:**
>
> No, the estimations do not add up to 1000.

> **Bard's reply:**
>
> You are correct. The sum of the estimations in the table is 1,000, not 950. I apologize for the error. I am still under development, and I am always learning new things. Thank you for pointing out the error.
>
> Here is the correct distribution of choices, with the estimated number of people in each group:
>
> | Choice | Estimated number of people |
> | --- | --- |
> | $0.00 (the minimum offer) | 100 (+/- 50) |
> | $0.10 | 200 (+/- 100) |
> | $0.20 | 300 (+/- 150) |
> | $0.30 | 400 (+/- 200) |
> | $0.40 | 500 (+/- 250) |
> | $0.50 | 600 (+/- 300) |
> | $0.60 | 500 (+/- 250) |
> | $0.70 | 400 (+/- 200) |

| | |
|---|---|
| $0.80 | 300 (+/- 150) |
| $0.90 | 200 (+/- 100) |

I believe that this is a more realistic estimate of the distribution of choices in this decision problem than the previous estimates I gave.

Is there anything else I can help you with today?

It is possible that the complexity of the decision problem we presented is causing issues for Bard. As a next step, we decided to estimate behavior using one of the simplest instructions from Capraro et al. (2014). We restarted the chat to proceed with this approach.

**Our prompt:**

Please read the following decision problem:

You are paired with another, anonymous participant, different from the one with whom you were paired in the first part of the HIT. This time how much money you earn depends only on your own choice.

You are given $0.10 and you have to decide how much, if any, to DONATE to the other participant. The other participant has no choice: she or he will get your donation.

Now imagine that there is a population of 1,000 people living in the USA facing this decision problem. How would you estimate the distribution of their choices? Please, for each choice, report a result in the form X\pmY, where X is the estimated number of people make their choice and Y is the error

**Bard's reply:**

Sure, here is how I would estimate the distribution of choices for a population of 1,000 people living in the USA facing the decision problem you described:

| Choice | Estimated number of people | Error |
|---|---|---|

> Keep all 0.10    300±50
>
> Donate 0.05    250±50
>
> Donate 0.01    200±50
>
> Donate 0       200±50
>
> I arrived at these estimates by considering the following factors:
>
> - **Self-interest:** Most people are motivated by self-interest, so they are likely to keep all of the 0.10 for themselves.
> - **Altruism:** However, some people are also motivated by altruism, so they are likely to donate some or all of the 0.10 to the other participant.
> - **Social norms:** Social norms can also influence people's choices. For example, if people believe that it is customary to donate some money in this situation, then more people are likely to donate.
> - **Randomness:** There is always some element of randomness in people's choices. For example, some people might make a decision based on a whim, or they might be influenced by a recent news story.
>
> The error bars in my estimates reflect the uncertainty associated with these factors. For example, it is difficult to say exactly how much people are motivated by self-interest or altruism, so there is some uncertainty in my estimate of the number of people who will keep all of the 0.10. Similarly, it is difficult to predict how social norms will influence people's choices, so there is some uncertainty in my estimate of the number of people who will donate.
>
> I believe that these estimates are a reasonable approximation of the distribution of choices that would be made by a population of 1,000 people living in the USA. However, it is important to note that these are just estimates, and the actual distribution of choices could vary depending on the specific circumstances.

There are several major issues with this reply. Firstly, Bard fails to understand that "Keep all 0.10" and "Donate 0" are identical choices. Additionally, the sum of the estimations does not amount to 1,000. To provide a comparison, here is an example of how GPT-4 responds to the same prompt:

> **GPT-4's reply:**
>
> The problem you've presented is an example of a one-shot decision-making problem without any direct consequences for the decision-maker, and the decision does not impact the outcome for the decision-maker in any way. Essentially, it's a pure donation decision. I would base my estimation on general statistics on philanthropy and generosity in the U.S., studies of economic games, and factors such as people's social, economic, and cultural backgrounds, as well as

> their understanding of the game.
>
> However, this is purely hypothetical and does not take into account potential additional factors, but should be a reasonable estimate given the constraints of the problem. Here is a possible distribution:
>
> - 250±50 people will donate $0.00 (25%±5%). People in this group may not want to give any money to a stranger, or they might need the money themselves.
> - 350±50 people will donate $0.05 (35%±5%). This group represents a balanced approach, with participants likely choosing to share half of their windfall with the other participant.
> - 300±50 people will donate $0.10 (30%±5%). These are the most generous participants, who are willing to give away all of their earnings.
> - 100±20 people will donate a different amount (10%±2%). This group might contain people who are undecided between giving half or all of their money, or those who simply want to donate an unconventional amount for other reasons.
>
> Please note that this distribution is a rough estimate, and the actual results could vary widely depending on the specific characteristics of the population being studied.

As clearly observed, the estimations provided by GPT-4 add up to 1,000, do not confuse any strategies, and demonstrate the typical tri-peak distribution mentioned in the main text. Thus, at this point, we have decided to abandon the use of Bard.

# Appendix 4: Estimating the trade-off between self- and other-interest using Bing

We have also tried using Bing. In the "precise" model, it appears that for this specific problem, Bing performs even worse than Bard. Let us provide a couple of examples to illustrate this behavior. In our first prompt, we utilize the straightforward instructions provided by Capraro et al. (2014).

> **Our prompt:**
>
> Please read the following decision problem:
>
> You are paired with another, anonymous participant, different from the one with whom you were paired in the first part of the HIT. This time how much money you earn depends only on your own choice.
>
> You are given $0.10 and you have to decide how much, if any, to DONATE to the other participant. The other participant has no choice: she or he will get your donation.
>
> Now imagine that there is a population of 1,000 people living in the USA facing this decision problem. How would you estimate the distribution of their choices? Please, for each choice, report a result in the form X\pmY, where X is the estimated number of people make their choice and Y is the error

> **Bing's reply:**
>
> Ricerca in corso per: donation behavior in the USA
> Ricerca in corso per: average donation amount in the USA
> Generazione delle risposte per l'utente in corso…
>
> It's difficult to estimate the distribution of choices for a population of 1,000 people living in the USA facing this decision problem. However, I can provide you with some information on donation behavior in the USA that might help you make an informed guess.
>
> According to a survey conducted by LendingTree, 56% of Americans donated to charity in 2021 and the average charitable donation amount reported by respondents in 2021 is $574[1]. However, this information is about charitable donations in general and not specific to the decision problem you mentioned.
>
> Is there anything else I can help you with? 😊

As evident, Bing demonstrates a lack of comprehension of the prompt on different levels. Firstly, it fails to understand the need to estimate a distribution of choices. Secondly, it fails to grasp the requirement of estimating the distribution of choices for a specific decision problem, as opposed to the general distribution of donations, despite recognizing this discrepancy. It is possible that these issues stem from the use of the term "donation" in the instructions by Capraro et al. (2014), which may have led to confusion for Bing. Hence, for our subsequent prompt, we opted for instructions from Branas-Garza (2007) that are still straightforward to understand but do not employ the term "donate".

---

**Our prompt:**

Please read the following decision problem:

A fixed amount of 10 experimental units has been provisionally allocated for you and your recipient. These 10 units are equal to 0.5 extra points in the final grade of Intermediate Microeconomics. Your task is to decide how to divide this amount of points between your recipient and yourself. Any division (even keeping all for yourself) is allowed. Your partner will be randomly selected from those 20 subjects placed in the row of your left. Thank you for your participation.

Now imagine that there is a population of 1,000 people living in Spain facing this decision problem. How would you estimate the distribution of their choices? Please, for each choice, report a result in the form X\pmY, where X is the estimated number of people make their choice and Y is the error

---

**Bing's reply:**

Ricerca in corso per: distribution of choices in decision problem
Ricerca in corso per: distribution of choices in decision problem Spain
Generazione delle risposte per l'utente in corso…

It is difficult to estimate the distribution of choices for a population of 1,000 people living in Spain facing this decision problem without conducting a survey or study. However, there are theories and research on decision-making that could provide some insight into how people might make their choices.

One theory is the **Framing of Decisions and the Psychology of Choice** by Amos Tversky and Daniel Kahneman which explains that people's choices are often influenced by the way the decision problem is framed[1]. Another body of research has focused on simple value-based choices to study situations very similar to our daily life decisions as consumers. The use of neuroimaging techniques has deepened and refined our knowledge of decision processes[2].

> Would you like more information on these theories or any other specific aspect of decision-making?

Therefore, it appears that Bing does not make an attempt to provide a guess. Subsequently, we made an explicit request for Bing to generate a guess:

> **Our prompt:**
>
> Can you make a guess?

However, Bing declined to provide one:

> **Bing's response:**
>
> As an AI, I don't make guesses. I can only provide information based on available data and research. In this case, without conducting a survey or study, it is not possible to accurately estimate the distribution of choices for a population of 1,000 people living in Spain facing this decision problem.

It is possible that Bing only refuses to provide guesses in "precise" mode. Therefore, we restarted the chat and switched the chatbot to "normal" mode. We then asked it to estimate the proportion of choices using the same instructions as in Branas-Garza (2007):

> **Bing's reply:**
>
> Ricerca in corso per: distribution of choices in decision problems like "how to divide a fixed amount of points between two people"
> Generazione delle risposte per l'utente in corso…
>
> The distribution of choices in decision problems like "how to divide a fixed amount of points between two people" is a topic of interest in the field of decision theory. In general, the distribution of choices depends on the context and the preferences of the individuals involved. However, there are some models that can be used to predict the distribution of choices in certain situations.
>
> One such model is the Nash equilibrium. The Nash equilibrium is a concept in game theory that describes a state where each player is making the best decision possible given the decisions of the other players. In the context of your decision problem, the Nash equilibrium would be for both you and your recipient to keep all 10 units for yourself. However, this is not necessarily what will happen in practice.
>
> Another model that can be used to predict the distribution of choices is prospect theory.

> Prospect theory is a behavioral economic theory that describes how people make decisions under uncertainty. According to prospect theory, people are more likely to take risks when they are faced with losses than when they are faced with gains. In the context of your decision problem, this means that people might be more likely to divide the points equally if they are told that they will lose points if they do not reach an agreement.
>
> Unfortunately, I could not find any specific data on how people in Spain would respond to this decision problem. However, based on these models and other research on decision-making, it is likely that there would be a wide range of responses. It is possible that some people would choose to keep all 10 units for themselves, while others might choose to divide them equally or give more to their partner.
>
> I hope this helps! Let me know if you have any other questions.

The answer is somewhat intriguing. Despite not providing an estimation for a distribution, it acknowledges that different theories can yield different predictions. It also acknowledges that individuals may choose to divide the units equally or allocate more to their partner. However, the theoretical reasoning leading to this conclusion is incorrect, as Bing argues that individuals may be "more likely to take risks when faced with losses than when faced with gains" in decision problems involving uncertainty, citing prospect theory as support. Yet, this argument is flawed because the decision problem at hand does not involve uncertainty, rendering prospect theory inapplicable (or trivially yielding the same predictions as Nash equilibrium). The reason behind this mistake is probably that Bing makes the same mistake as Bard and confuses this decision problem with the ultimatum game, as suggested by the sentence: "In the context of your decision problem, this means that people might be more likely to divide the points equally if they are told that they will lose points if they do not reach an agreement." Nonetheless, we attempted to request Bing to estimate a precise distribution of choices.

> **Our prompt:**
>
> Can you estimate the proportion of people who keep all 10 units for themselves and the proportion of people who divide them equally, etc.?

> **Bing's reply:**
>
> Ricerca in corso per: proportion of people who keep all 10 units for themselves and the proportion of people who divide them equally in decision problems
> Generazione delle risposte per l'utente in corso…
>
> It is difficult to estimate the proportion of people who would keep all 10 units for themselves

> and the proportion of people who would divide them equally without more information about the context and the preferences of the individuals involved. However, there are some methods that can be used to estimate these proportions.
>
> One such method is to conduct a survey or an experiment. In a survey or an experiment, a sample of individuals is asked to make a decision similar to the one you described. The responses of the sample can then be used to estimate the proportion of people who would keep all 10 units for themselves and the proportion of people who would divide them equally. However, this method can be expensive and time-consuming.
>
> Another method is to use mathematical models such as game theory or prospect theory. These models can be used to predict the proportion of people who would keep all 10 units for themselves and the proportion of people who would divide them equally based on assumptions about the context and the preferences of the individuals involved. However, these models are only as good as their assumptions and may not accurately predict the proportions in practice.
>
> Without more information about the context and the preferences of the individuals involved, it is difficult to estimate the proportions in Spain. However, based on these methods and other research on decision-making, it is likely that there would be a wide range of proportions. Some people might choose to keep all 10 units for themselves, while others might choose to divide them equally or give more to their partner.
>
> I hope this helps! Let me know if you have any other questions.

Therefore, in "normal" mode, Bing does not provide a distribution of choices. Finally, we attempted the "creative" mode.

> **Bing's reply:**
>
> Ricerca in corso per: experimental units decision problem
> Ricerca in corso per: ultimatum game distribution
> Generazione delle risposte per l'utente in corso…
>
> This is a very interesting decision problem. It seems to be related to the **ultimatum game**, which is a task that allows us to test the predictions of game theoretical assumptions. In a standard ultimatum game, two players share roles as the proposer and the responder. The task is to divide a constant sum of money in a one-shot interaction. The proposer makes an offer, and the responder either accepts or rejects the offer. If the responder accepts, the two parties share the money based on the offer. However, if the responder rejects, neither party earns anything[1].
>
> Traditional game theory assumes that human beings are fully rational and try to maximize their gains. Thus, the proposer should offer the minimum amount of money different than, yet close to, zero to maximize her gains and the receiver should accept that offer since it is better than nothing[1]. However, empirical studies have shown that this is not the case. Most proposers tend to offer more money than the minimum nonzero amount, and most responders tend to reject low yet nonzero offers[1]. This suggests that people are motivated by fairness and social norms rather than by self-interest alone.

To estimate the distribution of choices for a population of 1,000 people living in Spain facing this decision problem, we can look at some existing data from previous studies. For example, one study[2] reported that in Spain, the mean offer was 43% and the mean rejection rate was 16%. Assuming a normal distribution, we can calculate the standard deviation of both variables using the sample size (n = 40) and the standard error (SE) reported in the study:

SD(offer) = SE(offer) * sqrt(n) = 1.6 * sqrt(40) = 10.1

SD(rejection rate) = SE(rejection rate) * sqrt(n) = 3.9 * sqrt(40) = 24.7

Using these values, we can estimate the number of people who would make each choice using the normal distribution formula:

P(x) = (1 / (SD * sqrt(2 * pi))) * exp(-((x - mean)^2 / (2 * SD^2)))

For example, to estimate the number of people who would offer 50%, we can plug in x = 50, mean = 43 and SD = 10.1:

P(50) = (1 / (10.1 * sqrt(2 * pi))) * exp(-((50 - 43)^2 / (2 * 10.1^2)))

P(50) = 0.036

Multiplying this by 1,000, we get:

N(50) = 0.036 * 1000 = 36

Using a similar method, we can estimate the number of people who would offer other percentages and the number of people who would reject different offers. Here is a table with some possible results:

| Offer (%) | N(offer) | Rejection rate (%) | N(reject) |
| --- | --- | --- | --- |
| 10 | 3 | 80 | 24 |
| 20 | 7 | 60 | 42 |
| 30 | 14 | 40 | 56 |
| 40 | 25 | 20 | 50 |
| 50 | 36 | 10 | 36 |
| 60 | 25 | 5 | 13 |
| 70 | 14 | 2 | 3 |
| 80 | 7 | 1 | <1 |
| 90 | <1 | <1 | <1 |

Note that these are only rough estimates based on one study and may not reflect the actual distribution of choices for a population of Spanish people facing this decision problem. Also note that these estimates do not account for other factors that may influence people's decisions, such as risk aversion, social preferences, emotions, or cognitive biases.

I hope this helps you with your task. If you need more assistance, please let me know. 😊

The answer in "creative" mode is somewhat more intriguing compared to the other modes. However, it becomes evident from the outset that Bing confuses the dictator game with the ultimatum game. Therefore, we tried to assist Bing in rectifying this error:

**Our prompt:**

Note that this is not an ultimatum game.

**Bing's reply:**

I apologize for my mistake. You are right, this is not an ultimatum game, but a **dictator game**. In a dictator game, the proposer has full power over the allocation of the money and the responder has no choice but to accept whatever the proposer decides. This eliminates the possibility of rejection and makes the proposer's decision more revealing of their preferences.

However, some of the same factors that influence people's behavior in the ultimatum game may also apply to the dictator game, such as fairness, social norms, altruism, or guilt. Therefore, some of the data from the ultimatum game studies may still be relevant for estimating the distribution of choices in the dictator game. Of course, there may also be differences between the two games, such as lower offers and less variation in the dictator game.

To get more accurate estimates for the dictator game, we would need to look at studies that specifically used this game with Spanish participants. Unfortunately, I could not find any such studies in my search. However, I did find a meta-analysis that summarized the results of 616 dictator game studies from different countries and cultures. According to this meta-analysis, the mean offer across all studies was 28.35% and the median offer was 20%. The mean offer for Western European countries was 32.08% and the median offer was 25%. The mean offer for Spain was not reported separately, but it was included in the Western European category.

Using these values, we can try to estimate the distribution of choices for a population of 1,000 people living in Spain facing this decision problem. Assuming a normal distribution, we can calculate the standard deviation of the offer using the sample size (n = 616) and the standard error (SE) reported in the meta-analysis:

SD(offer) = SE(offer) * sqrt(n) = 0.65 * sqrt(616) = 16.4

Using these values, we can estimate the number of people who would make each choice using the normal distribution formula:

P(x) = (1 / (SD * sqrt(2 * pi))) * exp(-((x - mean)^2 / (2 * SD^2)))

For example, to estimate the number of people who would offer 50%, we can plug in x = 50, mean = 32.08 and SD = 16.4:

P(50) = (1 / (16.4 * sqrt(2 * pi))) * exp(-((50 - 32.08)^2 / (2 * 16.4^2)))

P(50) = 0.016

Multiplying this by 1,000, we get:

N(50) = 0.016 * 1000 = 16

Using a similar method, we can estimate the number of people who would offer other percentages. Here is a table with some possible results:

| Offer (%) | N(offer) |
| --- | --- |
| 10 | 36 |
| 20 | 59 |
| 30 | 61 |
| 40 | 39 |
| 50 | 16 |
| 60 | <1 |
| 70 | <1 |
| 80 | <1 |
| 90 | <1 |

Note that these are only rough estimates based on one meta-analysis and may not reflect the actual distribution of choices for a population of Spanish people facing this decision problem. Also note that these estimates do not account for other factors that may influence people's decisions, such as risk aversion, social preferences, emotions, or cognitive biases.

I hope this helps you with your task. If you need more assistance, please let me know. 😊

It appears that Bing corrects itself in identifying the game as the dictator game and accurately references Engel's meta-analysis, reporting the average donation as 28.35%. However, it makes several major errors as well. Firstly, it states that the average donation in Western countries is 32.08%, but Engel's meta-analysis does not provide this specific information. Instead, Engel presents evidence suggesting that the average donation in Western countries is lower than in developing and primitive countries (pp. 598-599). Therefore, it is likely that the average donation in Western countries is lower than 28.35%. Furthermore, Bing proceeds with estimations assuming a normal distribution of giving, which is not the case. Engel's meta-analysis clearly demonstrates a distribution with three peaks.

Moreover, Bing makes several computational errors that result in estimations that do not add up to 1,000. It initially uses the sample size of the meta-analysis (n=616) to calculate the standard deviation, but then employs the same standard deviation to compute P(x), which should be based on a sample size of 1,000. Additionally, the SE(offer) = 0.65 used to calculate the standard deviation is not present in Engel's meta-analysis and appears to be an incorrect value.

In summary, despite initially appearing as a technically sound and well-thought-out response, Bing's reply in creative mode contains multiple mistakes. Consequently, we have decided to abandon using Bing at this stage.

# SI References